\documentclass[conference]{IEEEtran}

\usepackage{bm} %provide \bm to bold math
\usepackage{booktabs} %for table
\usepackage{verbatim} %for comment

\usepackage{array}
\usepackage{cite}
\ifCLASSINFOpdf
  \usepackage[pdftex]{graphicx}
  \graphicspath{{./pic/}}
  \DeclareGraphicsExtensions{.pdf,.jpg,.png}
\else
\fi

\usepackage[boxed,ruled,linesnumbered]{algorithm2e}

\usepackage{fancyhdr}

\begin{document}

\title{DSLUT: An Asymmetric LUT and its Automatic Design Flow Based on Practical Functions}

\author{\IEEEauthorblockN{} 
\IEEEauthorblockA{\\ \\ \\} }

%for 4 authors
\author{\IEEEauthorblockN{Moucheng Yang*, Kaixiang Zhu*, Lingli Wang*, and Xuegong Zhou$^{\dagger}$}
\IEEEauthorblockA{*State Key Laboratory of ASIC and System, Fudan University, Shanghai, China \\ $^{\dagger}$Institute of Big Data, Fudan University, Shanghai, China \\ $^{\dagger}$zhouxg@fudan.edu.cn }}

\maketitle

\thispagestyle{fancy}
\fancyhead{} 
\lhead{} 
\lfoot{} 
\cfoot{} 
\rfoot{}

\begin{abstract}
The conventional LUT is redundant since practical functions in real-world benchmarks only occupy a small proportion of all the functions. For example, there are only 3881 out of more than $10^{14}$ NPN classes of 6-input functions occurring in the mapped netlists of the VTR8 and Koios benchmarks. Therefore, we propose a novel LUT-like architecture, named DSLUT, with asymmetric inputs and programmable bits to efficiently implement the practical functions in domain-specific benchmarks instead of all the functions. The compact structure of the MUX Tree in the conventional LUT is preserved, while fewer programmable bits are connected to the MUX Tree according to the bit assignment generated by the proposed algorithm. A 6-input DSLUT with 26 SRAM bits is generated for evaluation, which is based on the practical functions of 39 circuits from the VTR8 and Koios benchmarks. After the synthesis flow of ABC, the post-synthesis results show that the proposed DSLUT6 architecture reduces the number of levels by 10.98\% at a cost of 7.25\% area overhead compared to LUT5 architecture, while LUT6 reduces 15.16\% levels at a cost of 51.73\% more PLB area. After the full VTR flow, the post-implementation results show that the proposed DSLUT6 can provide performance improvement by 4.59\% over LUT5, close to 5.42\% of LUT6 over LUT5, causing less area overhead (6.81\% of DSLUT6 and 10.93\% of LUT6).\footnote{This work has been accepted for publication in the Proceedings of the International Conference on Field Programmable Technology (ICFPT), 2023. Personal use of this material is permitted. Permission from IEEE must be obtained for all other uses, in any current or future media, including reprinting/republishing this material for advertising or promotional purposes, creating new collective works, for resale or redistribution to servers or lists, or reuse of any copyrighted component of this work in other works.}
\end{abstract}


\section{Introduction} \label{sec:Intr}
Since its first introduction in the 1980s, the field programmable gate array (FPGA) has become an efficient alternative of implementation for digital circuits due to its flexibility and programmability. The SRAM-based look-up table (LUT) used as the programmable logic block (PLB) currently dominates the commercial FPGA architectures because of its complete functionality and EDA friendliness. However, it is quite expensive to raise the number of inputs (\emph{K}) of LUT due to the exponential area growth with \emph{K}\cite{ref_fpga_arch}. On the other hand, the LUT is redundant since practical functions in real-world benchmarks only occupy a small proportion of all the functions. For example, there are only 3881 out of more than $10^{14}$ NPN classes of 6-input functions in the mapped netlists of the VTR8\cite{ref_vtr8} and Koios\cite{ref_koios}  benchmarks.

In order to address the high cost of extension as well as the redundancy of functionality, we propose a novel LUT-like architecture, named DSLUT, with asymmetric inputs and programmable SRAM bits to efficiently implement the practical functions from the domain-specific benchmarks instead of all the functions. Asymmetric inputs prohibit arbitrary input permutation of DSLUT by changing programmable bits, which can be supported by programmable interconnect instead. The compact structure of the MUX Tree in the conventional LUT is preserved, while fewer programmable bits are connected to the MUX Tree according to the bit assignment generated by the proposed algorithm.

In this paper, A 6-input DSLUT with 26 SRAM bits is generated for evaluation, which is based on the practical functions of 39 circuits from the VTR8 and Koios benchmarks. After the synthesis flow of ABC\cite{ref_abc}, the post-synthesis results show that the proposed DSLUT6 architecture reduces the number of levels by 10.98\% at a cost of 7.25\% area overhead compared to LUT5 architecture, while LUT6 reduces 15.16\% levels at a cost of 51.73\% more PLB area. After the full VTR flow\cite{ref_vtr8}, the post-implementation results show that the proposed DSLUT6 achieves performance improvement by 4.59\% over LUT5, close to 5.42\% of LUT6 over LUT5, causing less area overhead (6.81\% of DSLUT6 and 10.93\% of LUT6). Both of the two experimental results indicate that the proposed DSLUT6 can raise the \emph{K} of PLB to improve performance at a lower cost of area. 

The major contributions of this paper are listed as follows\footnote{https://github.com/cmy1230/DSLUT.git}:

1) DSLUT: an asymmetric LUT-like architecture of PLB based on function-distribution analysis of the given benchmarks;

2) The generation of DSLUT by a combination of a heuristic algorithm and a hyper-parameter optimization algorithm; 

3) The SAT-based boolean matching of DSLUT;

4) The evaluation by the full EDA flow.

This paper is organized as follows. Section \ref{sec:Back} introduces some previous works and important methods or tools used. Section \ref{sec:Prop} describes the proposed DSLUT architecture in detail. Section \ref{sec:Expe} gives experimental results. Section \ref{sec:Conc} concludes this paper.

\section{Background} \label{sec:Back}
% 2.1 
\subsection{Previous Works on Exploration of PLB}
In the early years of integrated circuit technology, universal logic modules (ULM)\cite{ref_ulm} were introduced to facilitate modular synthesis techniques for logic networks, which can realize all the possible functions of a given number of inputs. For example, \cite{ref_ulm_example} designed a 5-input ULM by enumerating all the 3-input NPN classes, which can realize all the 3-input functions. However, the ULM required 36 inputs to implement all the 6-input functions\cite{ref_ulm_example2}. In order to reduce the number of I/O terminals, the prototype of LUT, called serially controlled ULM then, was first proposed in \cite{ref_ulm_lut}, whose 64 inputs were driven by constant 0 or 1 from internal shift registers and other 6 inputs by external I/O terminals. 

Many research efforts are devoted to improving the logic density and area efficiency of LUT. An extended LUT which contains a LUT and several logic gates is proposed in \cite{ref_elut} to raise the functionality. These two papers \cite{ref_lut_tree} and \cite{ref_sat_bm} both evaluate several generic LUT-based PLB architectures which consist of interconnected LUTs and logic gates. 

On the other hand, several non-LUT-like PLBs such as \cite{ref_hplb} and \cite{ref_hplb2} are designed by investigating the most frequently used functions in standard benchmarks. However, only 4-input functions are considered due to the limitation of manual design. In this paper, an automatic design flow is proposed to generate DSLUT by investigating functions with 6 or more inputs. 

Commercial FPGAs have different PLBs to raise area efficiency. The PLB proposed in \cite{ref_actel} consists of MUX4 and two logic gates connected. The Stratix 10 ALM\cite{ref_intel} has a fracturable PLB with 8 independent inputs which can implement two 5-input functions sharing 8 inputs or any 6-input functions. In Versal architecture from Xilinx\cite{ref_xilinx}, the LUT is enhanced to increase effective packing density and functionality so that two independent functions of up to 6 unique inputs can be packed in one PLB.

% 2.2
\subsection{Classification of Logic Function by NPN}
Logic function classification is the problem of grouping functions into equivalence classes. The most frequently used classification method is based on Negation-Permutation-Negation (NPN) equivalence. Two Boolean functions are NPN equivalent, if one can be obtained from the other by negating inputs, permuting inputs, and negating the output. 

In this paper, fast exact NPN classification\cite{ref_npn} is used to dramatically reduce complexities in function enumeration, generation of the function library, and boolean matching in technology mapping. The number of NPN classes of functions with different numbers of inputs is shown in Table \ref{tab:num_class}, where the NPN capacity is the maximum number of functions an NPN class can contain, and \emph{nPracticalClass} is the number of the practical NPN classes in mapped netlists of VTR8 and Koios benchmarks.

% Table generated by Excel2LaTeX from sheet 'num NPN class'
\begin{table}[htbp]
  \centering
  \caption{The number of NPN classes}
    \begin{tabular}{rrrrr}
    \toprule
    \multicolumn{1}{c}{nInputs} & \multicolumn{1}{c}{nFuncs} & \multicolumn{1}{c}{NPN capacity} & \multicolumn{1}{c}{nAllClass} & \multicolumn{1}{c}{nPracticalClass} \\
    \midrule
    K     & $2^{(2^K)}$ & $2^{(k+1)}*k!$ &       &  \\
    2     & 16    & 8     & 4     & 4 \\
    3     & 256   & 48    & 14    & 10 \\
    4     & 65536 & 768   & 222   & 150 \\
    5     & $4.3\times10^9$ & 7680  & 616126 & 1026 \\
    6     & $1.8\times10^{19}$ & 92160 & $\sim10^{14}$ & 3881 \\
    7     & $3.4\times10^{38}$ & 1290240 & $\sim10^{32}$ & 7843 \\
    8     & $1.2\times10^{77}$ & $2\times 10^7$ & $\sim10^{69}$ & 14309 \\
    \bottomrule
    \end{tabular}%
  \label{tab:num_class}%
\end{table}%

% 2.3
\subsection{Boolean Matching for General PLB} \label{subsec:bm}
If a PLB with incomplete functionality instead of the standard LUT is used in the FPGA, boolean matching has to be involved in the technology mapping flow, to determine whether a function can be implemented by the given PLB. There are three existing methods of boolean matching for FPGAs: function decomposition\cite{ref_lut_tree}, the pre-computed library\cite{ref_actel}, and boolean satisfiability\cite{ref_sat_bm}. The method based on function decomposition is fast but inflexible because a new boolean matching algorithm has to be designed for each new PLB. The method based on the pre-computed library has a high space complexity when the number of inputs is greater than 5 because all the functions that can be implemented by the given PLB are enumerated. The last SAT-based method is general for any PLBs, but very time-consuming. 

In this paper, we use an integrated SAT-based approach based on a pre-computed practical function library, which is similar to \cite{ref_dsd_map}, to map logic netlists into arbitrary single-output PLBs. In order to accelerate the SAT solving, the well-known paradigm of counterexample guided abstraction refinement (CEGAR)\cite{ref_cegar_qsat} is applied. The boolean matching in \cite{ref_dsd_map} only supports coarse-grained PLBs containing LUTs, MUXes, and elementary 2-input logic gates, while the SAT-based boolean matching implemented by us in Python is able to support fine-grained DSLUT.

% 2.4
\subsection{Automatic Hyper-parameter Optimization}
The generation of DSLUT can be formulated as a problem of hyper-parameter optimization, whose target is to maximize the coverage of a given DSLUT with the given number of SRAM bits. A meta-modeling approach to support automated hyper-parameter optimization is proposed in \cite{ref_hyperopt}, which has an open-source package of Python named \emph{hyperopt}.  

There are four important components of \emph{hyperopt}: the objective function to minimize, the space over which to search, the search algorithm to use, and the database in which to store all the point evaluations of the search. After the definitions of these four components, \emph{hyperopt} can find the best value of the scalar-valued, possibly-stochastic objective function over a set of possible arguments to that function.

% 2.5
%\subsection{Synthesis Flows by ABC}

\section{Proposed Architecture} \label{sec:Prop}
% 3.1
\subsection{DSLUT: Domain-Specific LUT} 
\begin{figure}[!t] %!t means top
\centering
\includegraphics[width=3in]{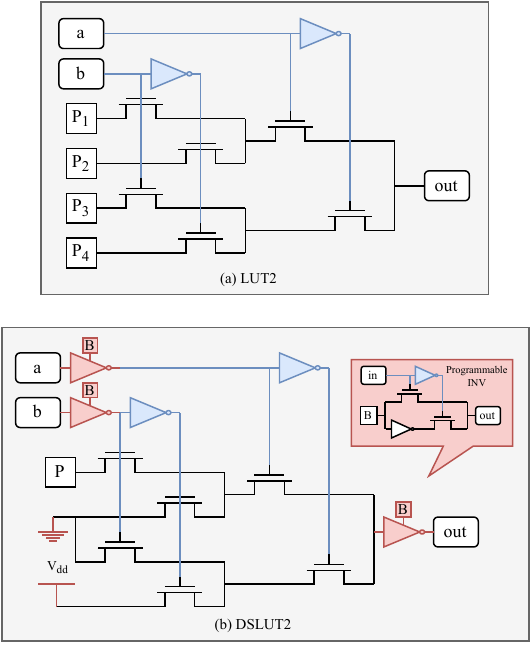}
\caption{LUT2 and DSLUT2}
\label{fig:lut2dslut2}
\end{figure}

In order to introduce the DSLUT, we first take LUT2 and DSLUT2 as examples, as shown in \figurename \ref{fig:lut2dslut2}. LUT2 can be seen as a ULM6, whose function $G$ is formulated as equation \ref{equ:lut2}. $\{P_i\}$ are four programmable SRAM bits, while $a$ and $b$ are two inputs. After $\{P_i\}$ is given, the function $G$ of LUT2 is then determined. There are 16 2-input functions in total, taking 4 bits to encode. However, if function classification by NPN is concerned, there are only 4 classes of 2-input functions, taking fewer bits to encode. 

\begin{equation}
\label{equ:lut2}
  G(a,b,P_1,P_2,P_3,P_4) = \overline{a}\overline{b}P_1 + a\overline{b}P_2 + \overline{a}bP_3 + abP_4
\end{equation}

As shown in \figurename \ref{fig:lut2dslut2}.(b), three programmable inverters (\emph{PINV}) are added to I/O terminals so that operations of I/O negation are supported. Therefore, we only need to consider one function for each NPN class. For example, we consider two function: $G(a,b,0,0,0,1)=ab$ and $G(a,b,1,0,0,1)=ab+\overline{a}\overline{b}$ representing two 2-input NPN classes. Only one programmable bit is required to encode these two NPN classes, that is $G(a,b,P,0,0,1)=H(a,b,P)=ab+P\overline{a}\overline{b}$, as shown in \figurename \ref{fig:lut2dslut2}.(b). Single-input and constant functions can be realized by bridging inputs, such as $H(a,a,0)=a$ and $H(a,\overline{a},0)=0$. So far, we still use four SRAM bits (the PINV takes one SRAM bit each) to encode four 2-input NPN classes. When the \emph{K} of DSLUT is increased, the reduction in the number of SRAM bits comes from 1) asymmetry of inputs; 2) input bridging, and 3) limited coverage of NPN classes. 

\begin{figure}[!t] %!t means top
\centering
\includegraphics[width=1.6in]{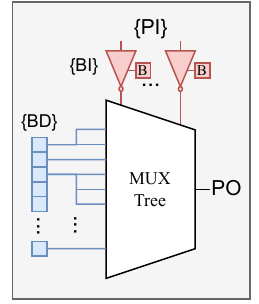}
\caption{Schematic of DSLUT}
\label{fig:dslut}
\end{figure}

Now we introduce the proposed asymmetric LUT-like PLB named DSLUT, as shown in \figurename \ref{fig:dslut}, where the compact structure of the MUX tree is preserved. $\{B_D\}$ are the actual existing SRAM bits. The connection pattern between $\{B_D\}$ and the data inputs of the MUX tree is called bit assignments. In order to decrease the delay overhead, the \emph{PINV} added to the output is removed. Therefore, the data input of the MUX tree cannot be driven by constant 0 or 1 so as to support output negation. Once the number of SRAM bits in $\{B_D\}$ and the bit assignment are fixed, the function of the DSLUT is determined.   

% 3.2 
\subsection{Library of Practical Functions} 
Since the algorithm of technology mapping, named as \emph{if} in ABC\cite{ref_abc}, includes the stage of cut enumeration\cite{ref_if}, it can be used to build a function library by harvesting functions in the given benchmarks. A novel representation of boolean functions in terms of their disjoint-support decomposition (DSD)\cite{ref_dsd} is used to make the function library more compact, because it is equivalent to NPN classification and supports fine-grained structure hashing. 

\begin{figure}[!t] %!t means top
\centering
\includegraphics[width=2.8in]{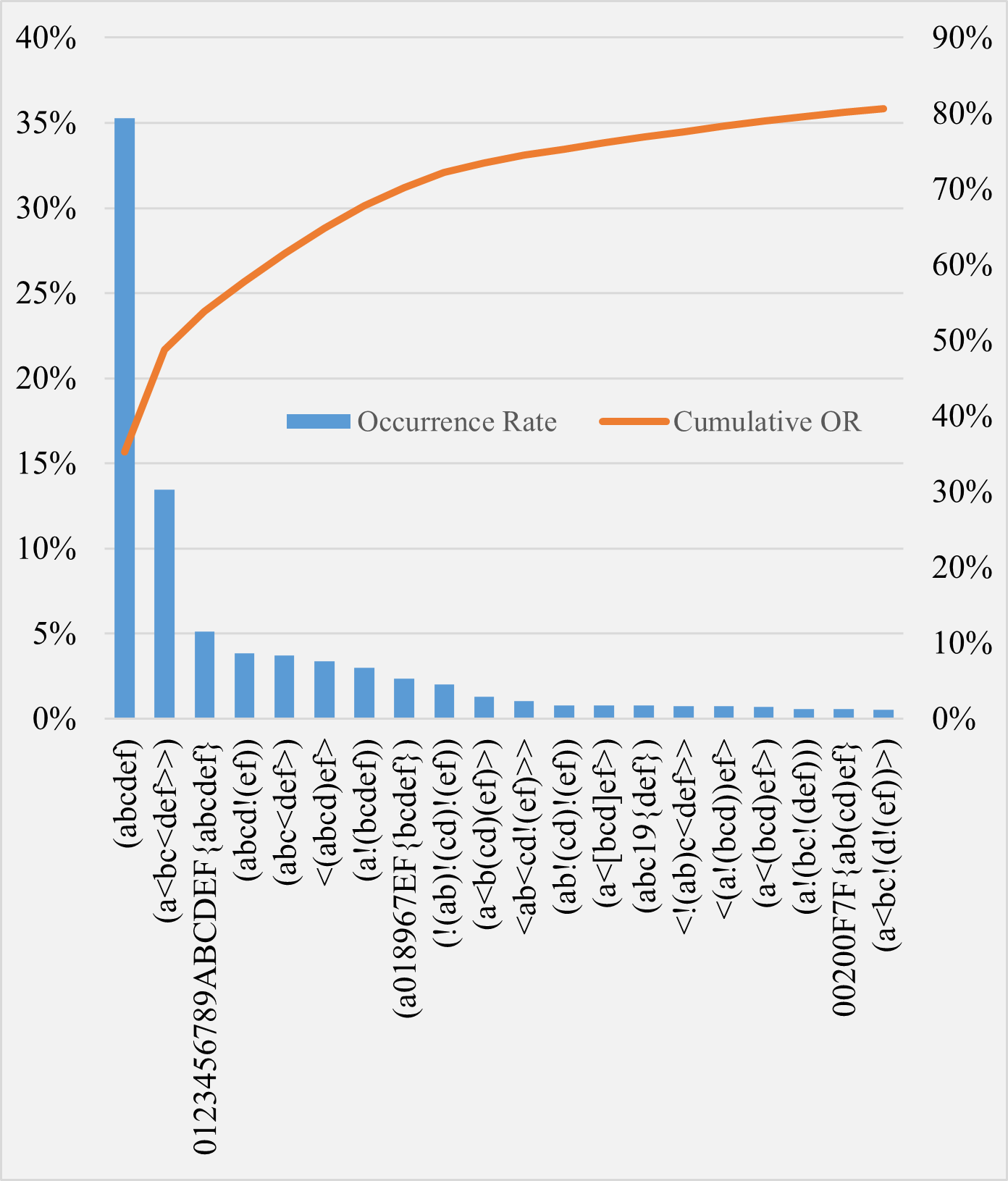}
\caption{Occurrence Rates of the TOP20 6-input Functions in VTR8 and Koios Benchmarks}
\label{fig:occur}
\end{figure}

Besides used for boolean matching as mentioned in Section \ref{subsec:bm}, the function library with additional statistical information can give an insight into function distribution. We define three kinds of occurrence of a function: \emph{nOccurEnum}, \emph{nOccurCutset} and \emph{nOccurCutBest}. For each function in the function library, its \emph{nOccurEnum} will count when it is enumerated in cut enumeration. Its \emph{nOccurCutset} will count when it is included in the cut set of an AIG node in \emph{If}, while its \emph{nOccurCutBest} will count when it is selected as a LUT in the mapped netlist. 

Occurrence counting is added into generation of the function library in \emph{if}. Occurrence rates are computed by dividing the \emph{nOccurCutBest} of a function by the sum of \emph{nOccurCutBest}s of all the functions. The occurrence rates of the top20 6-input functions in VTR8 and Koios benchmarks are shown in \figurename \ref{fig:occur}, which shows that the top 5\% (20 out of 3881) of 6-input functions account for 80\% of \emph{nOccurCutBest}s of all functions. The definition of expressions for boolean functions can be found in \cite{ref_dsd}.

\begin{comment}
% Table generated by Excel2LaTeX from sheet 'occurance'
\begin{table}[htbp]
  \centering
  \caption{The Occurrence Rate of Top20 6-input Practical Functions from VTR8 and KOIOS benchmarks}
    \begin{tabular}{crr}
    \toprule
    DSD String & \multicolumn{1}{l}{Occurrence Rate} & \multicolumn{1}{l}{Cumulative OR} \\
    \midrule
    (abcdef) & 35.255\% & 35.255\% \\
    (a\textless bc\textless def\textgreater \textgreater ) & 13.467\% & 48.721\% \\
    0123456789ABCDEF\{abcdef\} & 5.129\% & 53.850\% \\
    (abcd!(ef)) & 3.853\% & 57.703\% \\
    (abc\textless def\textgreater ) & 3.727\% & 61.430\% \\
    \textless (abcd)ef\textgreater  & 3.394\% & 64.824\% \\
    (a!(bcdef)) & 2.997\% & 67.821\% \\
    (a018967EF\{bcdef\}) & 2.345\% & 70.167\% \\
    (!(ab)!(cd)!(ef)) & 1.993\% & 72.160\% \\
    (a\textless b(cd)(ef)\textgreater ) & 1.308\% & 73.467\% \\
    \textless ab\textless cd!(ef)\textgreater \textgreater  & 1.051\% & 74.518\% \\
    (ab!(cd)!(ef)) & 0.782\% & 75.300\% \\
    (a\textless [bcd]ef\textgreater ) & 0.780\% & 76.080\% \\
    (abc19\{def\}) & 0.760\% & 76.841\% \\
    \textless !(ab)c\textless def\textgreater \textgreater  & 0.745\% & 77.586\% \\
    \textless (a!(bcd))ef\textgreater  & 0.723\% & 78.308\% \\
    (a\textless (bcd)ef\textgreater ) & 0.688\% & 78.996\% \\
    (a!(bc!(def))) & 0.579\% & 79.575\% \\
    00200F7F\{ab(cd)ef\} & 0.556\% & 80.131\% \\
    (a\textless bc!(d!(ef))\textgreater ) & 0.515\% & 80.646\% \\
    \bottomrule
    \end{tabular}%
  \label{tab:occur}%
\end{table}%
\end{comment}

% 3.3
\subsection{Generation of Bit Assignments}

\begin{algorithm}[!t]
\renewcommand{\algorithmcfname}{Algo}
\SetKwComment{Comment}{/*}{*/}
\SetKwComment{CommentOne}{//}{}
\SetKwInOut{Input}{Input} \SetKwInOut{Output}{Output}

\SetKwData{funclib}{funcLib}
\SetKwData{func}{func}
\SetKwData{funcs}{funcList}
\SetKwData{ninputs}{nInputs}
\SetKwData{nfuncs}{numTopFuncs}
\SetKwData{ttlist}{ttList}
\SetKwData{truthtable}{truthTable}
\SetKwData{tt}{tt}
\SetKwData{ttbest}{ttBest}
\SetKwData{ttcand}{ttCandidates}
\SetKwData{cost}{cost}
\SetKwData{costmin}{costMin}
\SetKwData{bainit}{bitAssignInit}
\SetKwData{bainitex}{bitAssignExt}
\SetKwData{ba}{bitAssign}
\SetKwData{bitmap}{bitMap}
\SetKwData{objective}{obj}
\SetKwData{searchspace}{space}
\SetKwData{algotouse}{algo}
\SetKwData{coverage}{coverage}

\SetKwFunction{listinit}{list}
\SetKwFunction{append}{append}
\SetKwFunction{remove}{remove}
\SetKwFunction{gettt}{getTt}
\SetKwFunction{loadlib}{loadLib}
\SetKwFunction{mergedup}{mergeDuplicate}
\SetKwFunction{sortfuncs}{sortByOccurrence}
\SetKwFunction{enumtt}{enumTtByNPN}
\SetKwFunction{computecost}{computeCost}
\SetKwFunction{gendslut}{genDslut}
\SetKwFunction{extendba}{extendBA}
\SetKwFunction{genspace}{genSpace}
\SetKwFunction{mixsuggest}{mixSuggest}
\SetKwFunction{fmin}{hyperopt.fmin}

\Input{nInputs, funcLib.dsd}
\Output{bitAssign}
  \funclib $\leftarrow$ \loadlib{"funcLib.dsd"}\;
  \funcs $\leftarrow$ [f for f in \funclib if f.nvars == nInputs]\;
  %\funcs $\leftarrow$ \mergedup{\funcs}\;
  \funcs $\leftarrow$ \sortfuncs{\funcs}\;
	\nfuncs $\leftarrow$ 8; \CommentOne{\small Predetermined}
  \ttlist $\leftarrow$ \listinit{}; \CommentOne{\small tt means truthTable}
  \For{i $\leftarrow$ 0 to (\nfuncs $-$ 1)}{
    \func $\leftarrow$ \funcs[\emph{i}]\;
    \eIf{i == 0}{
      \ttlist.\append{\func.\gettt{}}\;
    }{
      \ttcand $\leftarrow$ \enumtt{\func}\;
      \For{\tt in \ttcand}{
        \ttlist.\append{\tt}\;
        \cost $\leftarrow$ \computecost{\ttlist}\;
        \uIf{\cost $<$ \costmin}{
          \costmin $\leftarrow$ \cost \; 
          \ttbest $\leftarrow$ \tt \;
        }
        \ttlist.\remove{\tt}\;
      }
      \ttlist.\append{\ttbest}\;
    }
  }
  \bainit $\leftarrow$ \gendslut{\ttlist}\;
  \bainitex, \bitmap $\leftarrow$ \extendba{\bainit}\;
  \Comment{\small \emph{obj} computes the coverage rate of a DSLUT by boolean matching}
  \objective: \ba, \funcs $\rightarrow$ \coverage\;
  \searchspace $\leftarrow$ \genspace{\bainitex, \bitmap} \;
  \algotouse $\leftarrow$ \mixsuggest{(0.25,rand), (0.75,tpe)} \;
  \ba $\leftarrow$ \fmin{\objective, \searchspace, \algotouse}\;

\caption{The pseudo-code of DSLUT generation}
\label{alg:generation}
\end{algorithm}

\begin{figure}[!t] %!t means top
\centering
\includegraphics[width=2.8in]{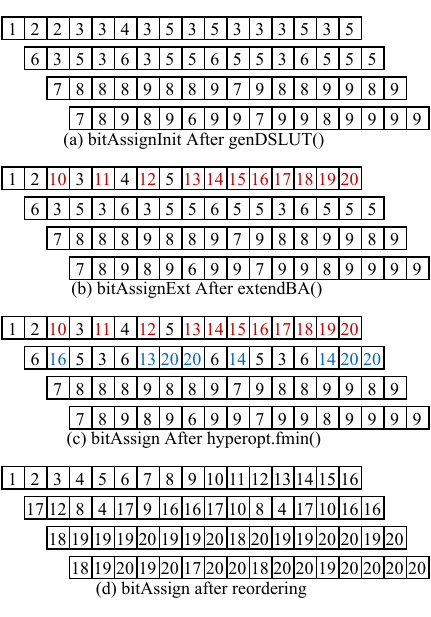}
\caption{An Example of Bit Assignment}
\label{fig:ba_example}
\end{figure}

The generation of bit assignments of a DSLUT is shown in Algorithm \ref{alg:generation}, which consists of two major stages: a heuristic algorithm for initialization and the hyper-parameter optimization by \emph{hyperopt}. Theorem 2.2.1 in \cite{ref_mfs} implies that a DSLUT with a given bit assignment can implement a function, if and only if for any two positions whose bits are different in the truth table of the function, the corresponding driving bits in $\{B_D\}$ according to the bit assignment are different. 

Taking DSLUT2 shown in \figurename \ref{fig:lut2dslut2}.(b) as example, the bit assignment $\{P_1, P_2, P_2, P_2\}$, resulting in $G(a,b,P_1,P_2,P_2,P_2)$ can implement the function $f$ with truth table (1,0,0,0), while the bit assignment $\{P_1, P_1, P_2, P_2\}$ can not. The reason is that the first two bits of the latter bit assignment are the same, while the first two bits of the truth table of $f$ are different, which makes it impossible to implement the function. 

Due to the additional PINVs, only one function of a NPN class needs to be considered. We can enumerate all the functions belonging to an NPN class to find the proper function as the representative, as shown in line 6 to line 22 in Algorithm \ref{alg:generation}. The minimum number of bits in the bit assignment for the given functions is computed in the function $computeCost()$ in line 14.  

In order to be compatible with LUT for small functions, the first 16 bits in the bit assignment are extended to be different from each other, resulting in a LUT4, as shown in line 24, while the changes caused by the extension are stored in \emph{bitMap}. New differences between bits in the bit assignment are introduced by \emph{hyperopt} within the limit of \emph{bitMap} to preserve the previous differences. 

The bit assignments in different stages of the generation of the proposed DSLUT6 are shown in \figurename \ref{fig:ba_example} as an example.

% 3.4
\subsection{Optimization of the MUX Tree} \label{subsec:muxtree}
\begin{figure}[!t] %!t means top
\centering
\includegraphics[width=2.4in]{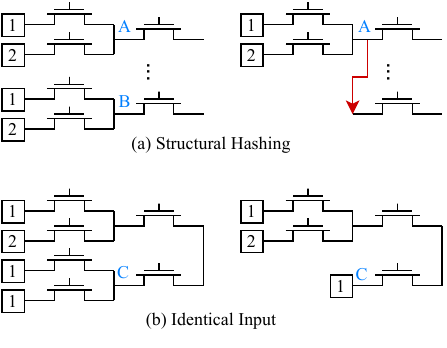}
\caption{Two Situations of MUX Tree Optimization}
\label{fig:mux_opt}
\end{figure}

Since the data inputs of the MUX tree are driven by fewer SRAM bits, the MUX tree itself can be optimized. There are two situations where the transistors in the MUX tree can be pruned, which are called structural hashing and identical inputs, as shown in \figurename \ref{fig:mux_opt}.

It is worth noting that when the transistors are removed due to identical inputs, as shown in \figurename \ref{fig:mux_opt}.(b), the delay between the removed input and the output in the optimized data path is reduced to zero, which will reduce the average delay of all the data paths of all the inputs.

% 3.5
\subsection{Delay and Area Modeling}

\begin{figure}[!t] %!t means top
\centering
\includegraphics[width=2.8in]{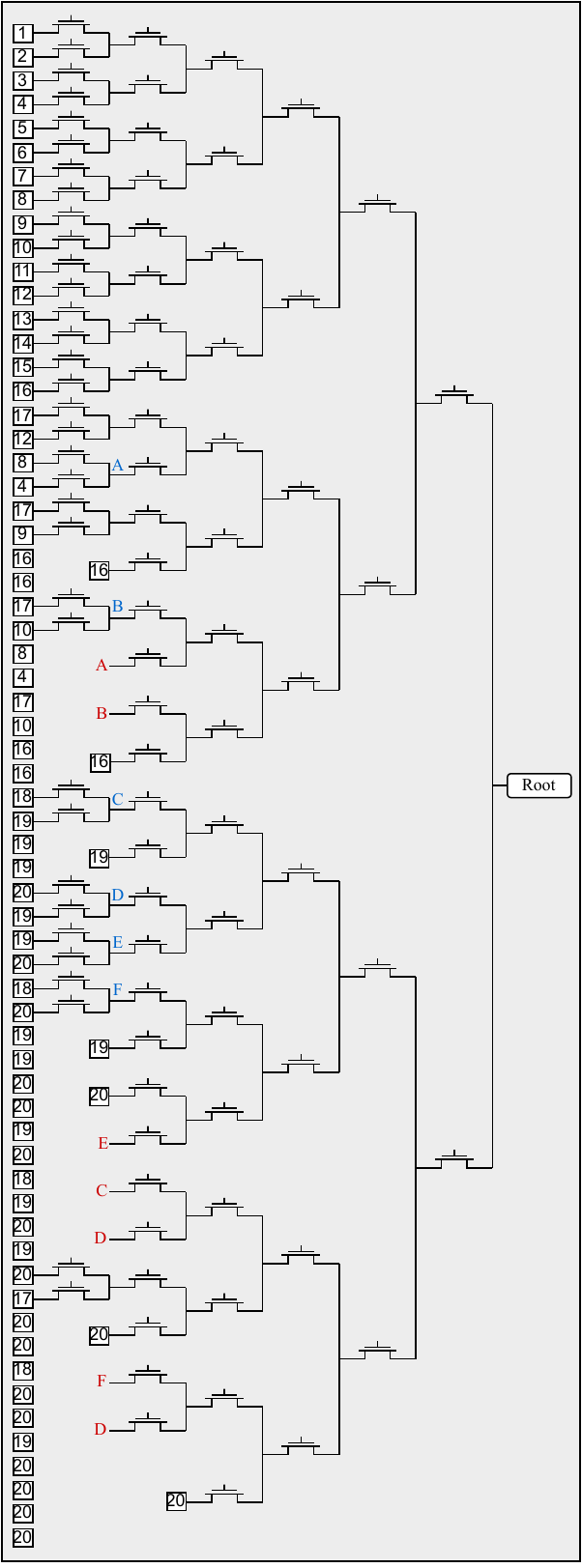}
\caption{MUX tree of the Proposed DSLUT6 with a given bit assignment}
\label{fig:golden_dslut}
\end{figure}

% Table generated by Excel2LaTeX from sheet 'model'
\begin{table}[htbp]
  \centering
  \caption{Modeling of DSLUT6 Based on COFFE\cite{ref_coffe} in 22nm}
    \begin{tabular}{lrrr}
    \toprule
    PLB area ($um^2$) & \multicolumn{1}{c}{LUT5} & \multicolumn{1}{c}{DSLUT6} & \multicolumn{1}{c}{LUT6} \\
    \midrule
    numSRAM & 32    & 26    & 64  \\
    total SRAM & 6.702  & 5.445  & 13.404  \\
    MUX tree & 3.747  & 5.799  & 7.481  \\
    input buffer & 2.901  & 4.730  & 3.525  \\
    other buffer & 3.555  & 2.828  & 3.751  \\
    \midrule
    total PLB & 16.905  & 18.802  & 28.161  \\
    \midrule
          &       &       &  \\
    \midrule
    CLB area ($um^2$) & \multicolumn{1}{c}{LUT5} & \multicolumn{1}{c}{DSLUT6} & \multicolumn{1}{c}{LUT6} \\
    \midrule
    10 PLB & 169.049  & 188.024  & 281.608  \\
    10 FF & 10.458  & 10.458  & 10.458  \\
    crossbar & 203.063 & 243.675 & 243.675 \\
    out mux & 21.872 & 21.872 & 21.872 \\
    carry & 44.856  & 44.856  & 44.856  \\
    \midrule
    total CLB & 449.298  & 510.066  & 602.469  \\
    \midrule
          &       &       &  \\
    \midrule
    delay ($ps$) & \multicolumn{1}{c}{LUT5} & \multicolumn{1}{c}{DSLUT6} & \multicolumn{1}{c}{LUT6} \\
    \midrule
    input[0] & 82  & 102  & 82  \\
	input[1] & 173  & 193  & 173  \\
	input[2] & 261 & 281 & 261 \\
	input[3] & 263 & 283 & 263 \\
	input[4] & 398  & 417  & 397  \\
	input[5] & N/A  & 418  & 398  \\
    \midrule
    average & 235.40   & 282.17   & 262.33  \\
    \bottomrule
    \end{tabular}%
  \label{tab:modeling}%
\end{table}%

The MUX tree and the bit assignment of the proposed DSLUT6 to be evaluated in Section \ref{sec:Expe} is shown in \figurename \ref{fig:golden_dslut}. 18 transistors are pruned due to identical inputs, while 14 transistors are pruned due to structural hashing. There are 94 transistors in the MUX tree of DSLUT6, while 62 and 126 transistors in the MUX tree of LUT5 and LUT6 respectively.

Since the DSLUT is quite similar to the LUT, the area models are based on the modeling results of fracturable LUT5 and LUT6 in 22nm from COFFE\cite{ref_coffe}, as shown in Table \ref{tab:modeling}, while the delay model is based on the delays in architecture description file \emph{k6\_frac\_N10\_frac\_chain\_mem32k\_40nm.xml}, which is used as the baseline architecture. 

\begin{figure}[!t] %!t means top
\centering
\includegraphics[width=2.4in]{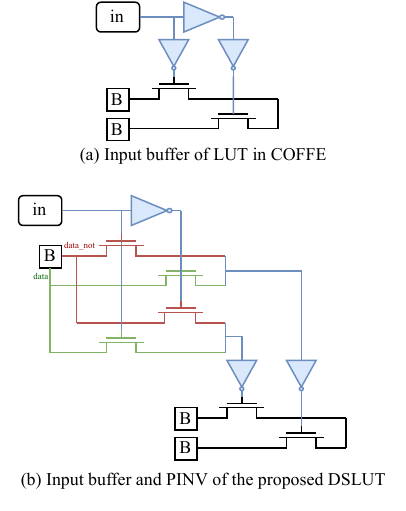}
\caption{The Input Buffer and Programmable INV}
\label{fig:input_buf}
\end{figure}

As shown in \figurename \ref{fig:input_buf}, the \emph{PINV}s added to the inputs of DSLUT6 cause the area overhead of 4 pass transistors and one SRAM bit. The \emph{other buffer} in Table \ref{tab:modeling} includes internal buffers and output buffers. The DSLUT6 is divided into two LUT3 in the fractured mode so that the output buffers are replaced by the internal buffers. Therefore, the area of the \emph{other buffer} is smaller than LUT5 and LUT6.

Furthermore, the difference in delay between DSLUT and LUT primarily stems from the additional PINVs added to the inputs. The delay overhead caused by the PINVs separately in the COFFE, which is 19.84ps, same as a 2-to-1 multiplexer with two inverters. Because VPR cannot support LUT rebalancing, the average of the delays of the LUTs or DSLUTs from all the inputs to the output is taken as the equivalent delay to get more stable results.

\section{Experimental Results} \label{sec:Expe}
% 4.1
\subsection{Experimental Methodology}
In order to demonstrate that the proposed DSLUT can extend the K of PLB to improve performance causing less area overhead than that of the LUT architecture, 39 benchmark circuits are synthesized and implemented in three PLB architectures including LUT5, DSLUT6 and LUT6, which share the same routing architecture. 

At first, the function coverage of the proposed DSLUT6 is to be evaluated. Then, the post-synthesis results of the ABC flow and the post-implementation results of the full VTR flow are discussed separately. 
The delays and areas of each architecture are averaged across all benchmark circuits, using the geometric average. The geometric average delay and area of each architecture is normalized to the LUT5 architecture for comparison.

% 4.2 
\subsection{Experimental Setups}
\subsubsection{Benchmarks}
39 circuits from the VTR8\cite{ref_vtr8} and Koios\cite{ref_koios} benchmark sets are used for the generation of the practical-function library and the architecture evaluations. 

\subsubsection{EDA Flow}
\begin{figure*}[!t] %!t means top
\centering
\includegraphics[width=6.6in]{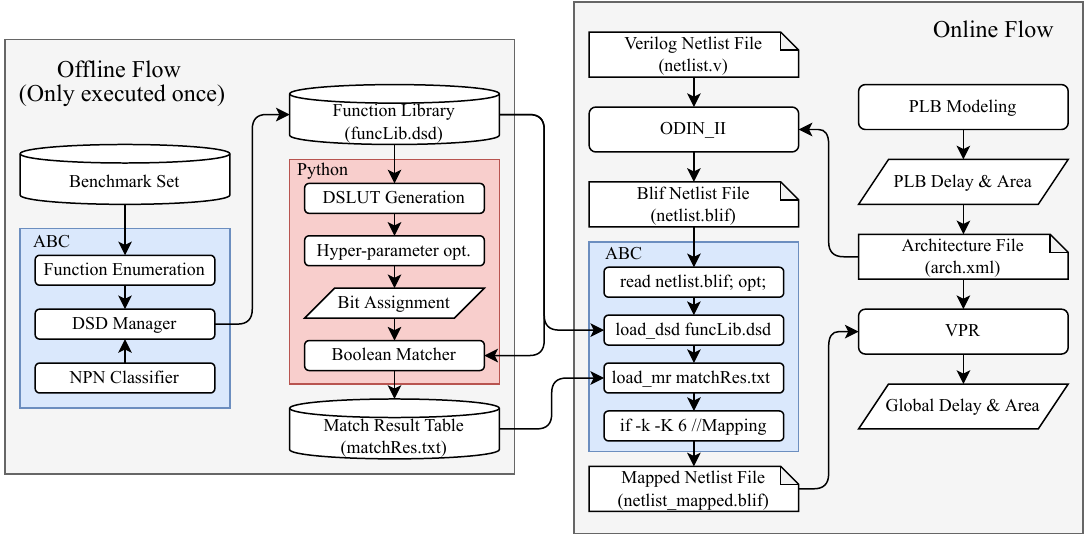}
\caption{Full EDA Flow}
\label{fig:EDA_flow}
\end{figure*}

As shown in \figurename \ref{fig:EDA_flow}, the EDA flow is divided into two major parts: the offline flow and the online flow. In the offline flow, the function library is built for boolean matching and DSLUT generation. In the online flow, the delay, and area information by PLB modeling is added to the architecture file first. When technology mapping, the function library and the matching results are loaded into ABC by corresponding instructions. Note that the instruction named \emph{load\_mr} is added by us specifically for the DSLUT. 

\subsubsection{Baseline Architectures}
The existing architecture file named \emph{k6\_frac\_N10\_frac\_chain\_mem32k\_40nm.xml}, which is included in the VTR directory, is used as the baseline architecture, with all the architectural parameters preserved for the LUT6 architecture to evaluate. For the LUT5 architecture, the PLB is replaced with a fracturable LUT5. For the DSLUT6 architecture, the PLB is replaced with the proposed DLSUT6, which can be fractured to two LUT3s. Since VPR cannot do LUT rebalancing, the average of the delays from all the inputs to the output is taken as the identical delays of inputs to get more stable results. The areas of the different PLBs and CLBs are shown in Table \ref{tab:modeling}. Moreover, the channel width of all the three architectures is 300.

% 4.3
\subsection{Coverage of Practical Functions}
% Table generated by Excel2LaTeX from sheet 'coverage'
\begin{table}[htbp]
  \centering
  \caption{The Coverage of the Proposed DSLUT6}
    \begin{tabular}{cc}
    \toprule
    nInputs & numClass \\
    \midrule
    3     & 10/10 \\
    4     & 150/150 \\
    5     & 910/1026 \\
    6     & 780/3881 \\
    \bottomrule
    \end{tabular}%
  \label{tab:coverage}%
\end{table}%

As shown in Table \ref{tab:coverage}, the proposed DSLUT6 can realize 780 out of 3881 NPN classes of 6-input practical functions in the library, using only 26 SRAM bits.

% 4.4 
\subsection{Post-Synthesis Results}
After the ABC flow, the detailed post-synthesis results of 39 benchmark circuits are shown in Table \ref{tab:post_syn}. The \emph{maxLevel} is the maximum number of PLBs in the critical path of the combinational parts of the benchmark circuit. The \emph{area} is computed by multiplying the \emph{nPLB} by the PLB area shown in Table \ref{tab:modeling}. The fracturability of the PLB is not taken into consideration.

The post-synthesis results show that the proposed DSLUT6 architecture reduces the number of levels by 10.98\% at a cost of 7.25\% area overhead compared to LUT5 architecture, while LUT6 reduces 15.16\% levels at a cost of 51.73\% more PLB area. In addition, the DSLUT6 architecture has the minimum delay-area product. 

% 4.5 
\subsection{Post-implementation Results}
After the full VTR flow, the detailed post-implementation results of 39 benchmark circuits are shown in Table \ref{tab:post_imp}. The \emph{delay} is extracted directly from the result file \emph{vpr.out}, while the \emph{area} is the product of the \emph{nCLB} and the CLB area shown in Table \ref{tab:modeling}. The area of the routing architecture is not considered since it is difficult to quantify the utilization of routing resources. 

The post-implementation results show that the proposed DSLUT6 can provide performance improvement by 4.59\% over LUT5, close to 5.42\% of LUT6 over LUT5, causing less logic area overhead (6.81\% of DSLUT6 and 10.93\% of LUT6). The delay-area product of the DSLUT6 architecture is better than that of the LUT6 architecture. 

% 5
\section{Conclusion} \label{sec:Conc}
An asymmetric LUT-like PLB architecture named DSLUT and its automatic design flow are proposed to address the high cost of extension as well as the redundancy of functionality in the conventional LUT. A library of practical functions harvested from a given benchmark set is built for the analysis of the functional distribution and boolean matching. An algorithm combining a heuristic initialization and hyper-parameter optimization is proposed to generate the bit assignment of DSLUT on the basis of the function library. 

The experimental results show that the proposed DSLUT6 containing 26 SRAM bits can implement 780 out of 3881 practical functions. Moreover, the FPGA architecture using DSLUT6 as PLB can achieve better performance by extending the \emph{K} of PLB with less logic area overhead.

% Table generated by Excel2LaTeX from sheet 'post_syn'
\begin{table*}[htbp]
  \renewcommand\arraystretch{0.83}
  \centering
  \caption{Post-Synthesis Results}
\scalebox{0.80}{
    \begin{tabular}{c|rrr|rrr|rrr}
    \toprule
          &       & \multicolumn{1}{c}{LUT5} &       &       & \multicolumn{1}{c}{DSLUT6} &       &       & \multicolumn{1}{c}{LUT6} &  \\
    netlist & \multicolumn{1}{c}{maxLevel} & \multicolumn{1}{c}{nPLB} & \multicolumn{1}{c|}{area ($um^2$)} & \multicolumn{1}{c}{maxLevel} & \multicolumn{1}{c}{nPLB} & \multicolumn{1}{c|}{area ($um^2$)} & \multicolumn{1}{c}{maxLevel} & \multicolumn{1}{c}{nPLB} & \multicolumn{1}{c}{area ($um^2$)} \\
    \midrule
    arm\_core & 23    & 15066 & 254691  & 19    & 14602 & 274547  & 19    & 13695 & 385665  \\
    attention\_layer & 9     & 22264 & 376373  & 8     & 21965 & 412986  & 7     & 21740 & 612220  \\
    bgm   & 25    & 56194 & 949960  & 21    & 54833 & 1030970  & 21    & 46738 & 1316189  \\
    blob\_merge & 6     & 7964  & 134631  & 6     & 7654  & 143911  & 5     & 7227  & 203520  \\
    bnn   & 3     & 82217 & 1389878  & 3     & 82086 & 1543381  & 3     & 82134 & 2312976  \\
    boundtop & 12    & 3312  & 55989  & 9     & 3119  & 58643  & 10    & 2982  & 83976  \\
    clstm\_like.large & 5     & 864599 & 14616046  & 4     & 849469 & 15971716  & 4     & 831080 & 23404044  \\
    clstm\_like.medium & 5     & 585804 & 9903017  & 4     & 576274 & 10835104  & 4     & 563577 & 15870892  \\
    clstm\_like.small & 5     & 306707 & 5184882  & 4     & 302644 & 5690312  & 4     & 296007 & 8335853  \\
    conv\_layer & 5     & 26061 & 440561  & 5     & 25872 & 486445  & 4     & 25514 & 718500  \\
    conv\_layer\_hls & 8     & 9976  & 168644  & 8     & 9291  & 174689  & 7     & 9410  & 264995  \\
    dla\_like.medium & 6     & 278077 & 4700892  & 5     & 276927 & 5206781  & 5     & 268297 & 7555512  \\
    dla\_like.small & 6     & 117242 & 1981976  & 5     & 115531 & 2172214  & 5     & 111169 & 3130630  \\
    eltwise\_layer & 6     & 15556 & 262974  & 5     & 14888 & 279924  & 5     & 14528 & 409123  \\
    gemm\_layer & 10    & 424464 & 7175564  & 9     & 386424 & 7265544  & 9     & 372094 & 10478539  \\
    lstm  & 8     & 106762 & 1804812  & 7     & 106312 & 1998878  & 7     & 105829 & 2980250  \\
    LU32PEEng & 129   & 97424 & 1646953  & 113   & 90927 & 1709609  & 108   & 84037 & 2366566  \\
    LU64PEEng & 131   & 188219 & 3181842  & 112   & 174859 & 3287699  & 109   & 161835 & 4557435  \\
    LU8PEEng & 130   & 28827 & 487320  & 113   & 27351 & 514254  & 107   & 25443 & 716500  \\
    mcml  & 32    & 105205 & 1778491  & 27    & 104012 & 1955634  & 26    & 101070 & 2846232  \\
    mkDelayWorker32B & 9     & 6473  & 109426  & 8     & 6103  & 114749  & 7     & 5992  & 168741  \\
    mkPktMerge & 3     & 389   & 6576  & 3     & 387   & 7276  & 3     & 381   & 10729  \\
    mkSMAdapter4B & 6     & 2352  & 39761  & 6     & 2246  & 42229  & 5     & 2078  & 58519  \\
    or1200 & 10    & 3362  & 56835  & 8     & 3072  & 57760  & 8     & 2791  & 78597  \\
    raygentop & 4     & 2566  & 43378  & 4     & 2474  & 46516  & 4     & 2292  & 64545  \\
    reduction\_layer & 7     & 11979 & 202505  & 6     & 12101 & 227523  & 6     & 9097  & 256181  \\
    robot\_rl & 17    & 22996 & 388747  & 15    & 21705 & 408097  & 14    & 20701 & 582961  \\
    sha   & 4     & 2094  & 35399  & 3     & 2506  & 47118  & 3     & 1620  & 45621  \\
    softmax & 12    & 23733 & 401206  & 10    & 21470 & 403679  & 10    & 20975 & 590677  \\
    spmv  & 8     & 7138  & 120668  & 7     & 6179  & 116178  & 6     & 5732  & 161419  \\
    spree & 19    & 1055  & 17835  & 16    & 1021  & 19197  & 15    & 931   & 26218  \\
    stereovision0 & 6     & 14365 & 242840  & 6     & 14344 & 269696  & 5     & 14152 & 398534  \\
    stereovision1 & 3     & 13114 & 221692  & 3     & 12601 & 236924  & 3     & 12490 & 351731  \\
    stereovision2 & 3     & 19452 & 328836  & 3     & 19402 & 364796  & 3     & 19391 & 546070  \\
    stereovision3 & 5     & 242   & 4091  & 5     & 225   & 4230  & 4     & 208   & 5857  \\
    tiny\_darknet\_like.medium & 8     & 292287 & 4941112  & 8     & 274039 & 5152481  & 7     & 266097 & 7493558  \\
    tiny\_darknet\_like.small & 8     & 102991 & 1741063  & 7     & 92378 & 1736891  & 7     & 89027 & 2507089  \\
    tpu\_like.medium & 6     & 105025 & 1775448  & 5     & 97872 & 1840189  & 5     & 97656 & 2750091  \\
    tpu\_like.small & 5     & 27147 & 458920  & 5     & 26867 & 505153  & 5     & 26470 & 745422  \\
    \midrule
    geomean & 9.043  & 24140.78 & 408100.0  & 8.050  & 23279.44 & 437700.1  & 7.672  & 21987.79 & 619198.1  \\
    percentage & 100\% & 100\% & 100\% & 89.02\% & 96.43\% & 107.25\% & 84.84\% & 91.08\% & 151.73\% \\
    \midrule
    delay-area product & 3690560 & 100\% &       & 3523559 & 95.47\% &       & 4750744 & 128.73\% &  \\
    \bottomrule
    \end{tabular}%
}
  \label{tab:post_syn}%
\end{table*}%

% Table generated by Excel2LaTeX from sheet 'post_imp'
\begin{table*}[htbp]
  \renewcommand\arraystretch{0.83}
  \centering
  \caption{Post-Implementation Results: Critical Path Delay and Logic Area}
\scalebox{0.80}{
    \begin{tabular}{c|rrr|rrr|rrr}
    \toprule
          &       & \multicolumn{1}{c}{LUT5} &       &       & \multicolumn{1}{c}{DSLUT6} &       &       & \multicolumn{1}{c}{LUT6} &  \\
    netlist & \multicolumn{1}{c}{delay ($ns$)} & \multicolumn{1}{c}{nCLB} & \multicolumn{1}{c|}{area ($um^2$)} & \multicolumn{1}{c}{delay ($ns$)} & \multicolumn{1}{c}{nCLB} & \multicolumn{1}{c|}{area ($um^2$)} & \multicolumn{1}{c}{delay ($ns$)} & \multicolumn{1}{c}{nCLB} & \multicolumn{1}{c}{area ($um^2$)} \\
    \midrule
    arm\_core & 19.774  & 1332  & 598464.9  & 20.024  & 1172  & 596413.2  & 19.035  & 1094  & 659101.1  \\
    attention\_layer & 9.026  & 1715  & 770546.1  & 8.724  & 1586  & 807091.6  & 8.255  & 1489  & 897076.3  \\
    bgm   & 27.434  & 5055  & 2271201.4  & 27.509  & 5278  & 2685895.0  & 26.552  & 4355  & 2623752.5  \\
    blob\_merge & 15.922  & 707   & 317653.7  & 14.330  & 663   & 337390.8  & 14.835  & 618   & 372325.8  \\
    bnn   & 8.776  & 7514  & 3376025.2  & 8.850  & 6411  & 3262461.7  & 9.591  & 6382  & 3844957.2  \\
    boundtop & 8.254  & 278   & 124904.8  & 6.901  & 272   & 138416.7  & 7.413  & 244   & 147002.4  \\
    clstm\_like.large & 19.074  & 82231 & 36946223.8  & 20.162  & 76863 & 39114427.8  & 15.478  & 60379 & 36376475.8  \\
    clstm\_like.medium & 20.548  & 55337 & 24862803.4  & 14.737  & 51882 & 26401971.6  & 13.449  & 40766 & 24560251.3  \\
    clstm\_like.small & 12.972  & 28582 & 12841835.4  & 10.624  & 26911 & 13694604.2  & 9.506  & 21172 & 12755473.7  \\
    conv\_layer & 5.907  & 2174  & 976773.9  & 5.640  & 1962  & 998432.4  & 6.224  & 1806  & 1088059.0  \\
    conv\_layer\_hls & 14.343  & 2157  & 969135.8  & 13.928  & 2069  & 1052883.1  & 13.640  & 1998  & 1203733.1  \\
    dla\_like.medium & 9.811  & 18754 & 8426134.7  & 9.448  & 18266 & 9295293.4  & 10.349  & 17385 & 10473923.6  \\
    dla\_like.small & 8.850  & 8145  & 3659532.2  & 10.461  & 8058  & 4100595.3  & 9.092  & 7579  & 4566112.6  \\
    eltwise\_layer & 5.415  & 1245  & 559376.0  & 5.076  & 1209  & 615242.0  & 5.432  & 1098  & 661511.0  \\
    gemm\_layer & 16.011  & 38540 & 17315944.9  & 14.450  & 35527 & 18079157.4  & 15.796  & 31818 & 19169358.6  \\
    lstm  & 9.100  & 8481  & 3810496.3  & 9.398  & 6850  & 3485862.3  & 9.382  & 6364  & 3834112.7  \\
    LU32PEEng & 88.817  & 8873  & 3986621.2  & 79.887  & 8466  & 4308220.4  & 80.329  & 7610  & 4584789.1  \\
    LU64PEEng & 87.064  & 17140 & 7700967.7  & 78.800  & 16188 & 8237830.4  & 80.517  & 14649 & 8825568.4  \\
    LU8PEEng & 88.325  & 2625  & 1179407.3  & 83.682  & 2508  & 1276283.6  & 81.947  & 2254  & 1357965.1  \\
    mcml  & 49.664  & 9617  & 4320898.9  & 40.374  & 9413  & 4790134.5  & 42.781  & 6870  & 4138962.0  \\
    mkDelayWorker32B & 8.687  & 569   & 255650.6  & 8.591  & 550   & 279886.8  & 8.776  & 524   & 315693.8  \\
    mkPktMerge & 4.088  & 28    & 12580.3  & 3.775  & 22    & 11195.5  & 4.202  & 21    & 12651.8  \\
    mkSMAdapter4B & 6.292  & 238   & 106932.9  & 6.049  & 224   & 113990.2  & 5.431  & 204   & 122903.7  \\
    or1200 & 8.482  & 323   & 145123.3  & 8.302  & 300   & 152665.5  & 8.718  & 267   & 160859.2  \\
    raygentop & 5.159  & 225   & 101092.1  & 4.933  & 226   & 115008.0  & 4.732  & 183   & 110251.8  \\
    reduction\_layer & 7.455  & 1197  & 537809.7  & 7.452  & 1272  & 647301.7  & 6.705  & 986   & 594034.4  \\
    robot\_rl & 11.516  & 1968  & 884218.5  & 11.032  & 1951  & 992834.6  & 11.142  & 1617  & 974192.4  \\
    sha   & 10.587  & 215   & 96599.1  & 10.180  & 259   & 131801.2  & 9.829  & 166   & 100009.9  \\
    softmax & 10.795  & 1718  & 771894.0  & 9.174  & 1662  & 845766.9  & 9.640  & 1403  & 845264.0  \\
    spmv  & 6.769  & 846   & 380106.1  & 6.907  & 679   & 345532.9  & 6.987  & 620   & 373530.8  \\
    spree & 11.315  & 97    & 43581.9  & 12.439  & 90    & 45799.7  & 10.957  & 73    & 43980.2  \\
    stereovision0 & 3.834  & 840   & 377410.3  & 3.857  & 829   & 421865.7  & 3.555  & 801   & 482577.7  \\
    stereovision1 & 5.490  & 904   & 406165.4  & 5.374  & 824   & 419321.2  & 5.417  & 796   & 479565.3  \\
    stereovision2 & 15.592  & 1828  & 821316.7  & 15.291  & 1670  & 849838.0  & 15.109  & 1660  & 1000098.5  \\
    stereovision3 & 2.665  & 20    & 8986.0  & 2.785  & 20    & 10177.7  & 2.592  & 13    & 7832.1  \\
    tiny\_darknet\_like.medium & 20.333  & 27178 & 12211021.0  & 16.854  & 24268 & 12349621.2  & 18.916  & 22143 & 13340471.1  \\
    tiny\_darknet\_like.small & 19.760  & 9986  & 4486689.8  & 20.112  & 8706  & 4430352.8  & 21.383  & 8030  & 4837826.1  \\
    tpu\_like.medium & 10.486  & 6559  & 2946945.6  & 10.220  & 6068  & 3087914.2  & 10.411  & 5639  & 3397322.7  \\
    tpu\_like.small & 7.660  & 1773  & 796605.4  & 7.666  & 1789  & 910395.3  & 7.400  & 1590  & 957925.7  \\
    \midrule
    geomean & 11.987  & 2108  & 947145.5  & 11.436  & 1988  & 1011654.6  & 11.337  & 1744  & 1050633.7  \\
    percentage & 100\% & 100\% & 100\% & 95.41\% & 94.30\% & 106.81\% & 94.58\% & 82.72\% & 110.93\% \\
    \midrule
    delay-area product & 11353015.9 & 100\% &       & 11569750 & 101.91\% &       & 11911261 & 104.92\% &  \\
    \bottomrule
    \end{tabular}%
}
  \label{tab:post_imp}%
\end{table*}%

%It's a test for cite:\cite{IEEEexample:confwithpapertype}

%\section*{Acknowledgment}
%The authors would like to thank???


\bibliographystyle{IEEEtran}
\bibliography{paper}

\begin{thebibliography}{10}
\providecommand{\url}[1]{#1}
\csname url@samestyle\endcsname
\providecommand{\newblock}{\relax}
\providecommand{\bibinfo}[2]{#2}
\providecommand{\BIBentrySTDinterwordspacing}{\spaceskip=0pt\relax}
\providecommand{\BIBentryALTinterwordstretchfactor}{4}
\providecommand{\BIBentryALTinterwordspacing}{\spaceskip=\fontdimen2\font plus
\BIBentryALTinterwordstretchfactor\fontdimen3\font minus
  \fontdimen4\font\relax}
\providecommand{\BIBforeignlanguage}[2]{{%
\expandafter\ifx\csname l@#1\endcsname\relax
\typeout{** WARNING: IEEEtran.bst: No hyphenation pattern has been}%
\typeout{** loaded for the language `#1'. Using the pattern for}%
\typeout{** the default language instead.}%
\else
\language=\csname l@#1\endcsname
\fi
#2}}
\providecommand{\BIBdecl}{\relax}
\BIBdecl

\bibitem{ref_fpga_arch}
J.~Rose, A.~El~Gamal, and A.~Vincentelli, ``Architecture of field-programmable
  gate arrays,'' \emph{Proceedings of the IEEE}, vol.~81, pp. 1013 -- 1029, 08
  1993.

\bibitem{ref_vtr8}
\BIBentryALTinterwordspacing
K.~E. Murray, O.~Petelin, S.~Zhong, J.~M. Wang, M.~Eldafrawy, J.-P. Legault,
  E.~Sha, A.~G. Graham, J.~Wu, M.~J.~P. Walker, H.~Zeng, P.~Patros, J.~Luu,
  K.~B. Kent, and V.~Betz, ``Vtr 8: High-performance cad and customizable fpga
  architecture modelling,'' \emph{ACM Trans. Reconfigurable Technol. Syst.},
  vol.~13, no.~2, jun 2020. [Online]. Available:
  \url{https://doi.org/10.1145/3388617}
\BIBentrySTDinterwordspacing

\bibitem{ref_koios}
A.~Arora, A.~Boutros, D.~Rauch, A.~Rajen, A.~Borda, S.~A. Damghani, S.~Mehta,
  S.~Kate, P.~Patel, K.~B. Kent, V.~Betz, and L.~K. John, ``Koios: A deep
  learning benchmark suite for fpga architecture and cad research,'' in
  \emph{2021 31st International Conference on Field-Programmable Logic and
  Applications (FPL)}, 2021, pp. 355--362.

\bibitem{ref_abc}
R.~Brayton and A.~Mishchenko, ``Abc: An academic industrial-strength
  verification tool,'' in \emph{Computer Aided Verification}, T.~Touili,
  B.~Cook, and P.~Jackson, Eds.\hskip 1em plus 0.5em minus 0.4em\relax Berlin,
  Heidelberg: Springer Berlin Heidelberg, 2010, pp. 24--40.

\bibitem{ref_ulm}
\BIBentryALTinterwordspacing
H.~S. STONE, ``Chapter vi - universal logic modules,'' in \emph{Recent
  Developments in Switching Theory}, A.~Mukhopadhyay, Ed.\hskip 1em plus 0.5em
  minus 0.4em\relax Academic Press, 1971, pp. 229--254. [Online]. Available:
  \url{https://www.sciencedirect.com/science/article/pii/B9780125098502500117}
\BIBentrySTDinterwordspacing

\bibitem{ref_ulm_example}
X.~Chen and X.~Wu, ``\BIBforeignlanguage{English}{Derivation of universal logic
  modules by algebraic means},'' \emph{\BIBforeignlanguage{English}{IEE
  Proceedings E (Computers and Digital Techniques)}}, vol. 128, pp.
  205--211(6), September 1981.

\bibitem{ref_ulm_example2}
F.~Preparata, ``On the design of universal boolean functions,'' \emph{IEEE
  Transactions on Computers}, vol. C-20, no.~4, pp. 418--423, 1971.

\bibitem{ref_ulm_lut}
S.~Yau and C.~Tang, ``Universal logic modules and their applications,''
  \emph{IEEE Transactions on Computers}, vol. C-19, no.~2, pp. 141--149, 1970.

\bibitem{ref_elut}
J.~H. Anderson, Q.~Wang, and C.~Ravishankar, ``Raising fpga logic density
  through synthesis-inspired architecture,'' \emph{IEEE Transactions on Very
  Large Scale Integration (VLSI) Systems}, vol.~20, no.~3, pp. 537--550, 2012.

\bibitem{ref_lut_tree}
J.~Cong and Y.-Y. Hwang, ``Boolean matching for lut-based logic blocks with
  applications to architecture evaluation and technology mapping,'' \emph{IEEE
  Transactions on Computer-Aided Design of Integrated Circuits and Systems},
  vol.~20, no.~9, pp. 1077--1090, 2001.

\bibitem{ref_sat_bm}
A.~C. Ling, D.~P. Singh, and S.~D. Brown, ``Fpga plb architecture evaluation
  and area optimization techniques using boolean satisfiability,'' \emph{IEEE
  Transactions on Computer-Aided Design of Integrated Circuits and Systems},
  vol.~26, no.~7, pp. 1196--1210, 2007.

\bibitem{ref_hplb}
Y.~Okamoto, Y.~Ichinomiya, M.~Amagasaki, M.~Iida, and T.~Sueyoshi, ``Cogre: A
  configuration memory reduced reconfigurable logic cell architecture for area
  minimization,'' in \emph{2010 International Conference on Field Programmable
  Logic and Applications}.\hskip 1em plus 0.5em minus 0.4em\relax IEEE, 2010,
  pp. 304--309.

\bibitem{ref_hplb2}
I.~Ahmadpour, B.~Khaleghi, and H.~Asadi, ``An efficient reconfigurable
  architecture by characterizing most frequent logic functions,'' in \emph{2015
  25th International Conference on Field Programmable Logic and Applications
  (FPL)}.\hskip 1em plus 0.5em minus 0.4em\relax IEEE, 2015, pp. 1--6.

\bibitem{ref_actel}
A.~Kennings, K.~Vorwerk, A.~Kundu, V.~Pevzner, and A.~Fox, ``Fpga technology
  mapping with encoded libraries and staged priority cuts,'' \emph{ACM
  Transactions on Reconfigurable Technology and Systems (TRETS)}, vol.~4,
  no.~4, pp. 1--17, 2011.

\bibitem{ref_intel}
D.~Lewis, G.~Chiu, J.~Chromczak, D.~Galloway, B.~Gamsa, V.~Manohararajah,
  I.~Milton, T.~Vanderhoek, and J.~Van~Dyken, ``The stratix™ 10 highly
  pipelined fpga architecture,'' in \emph{Proceedings of the 2016 ACM/SIGDA
  International Symposium on Field-Programmable Gate Arrays}, 2016, pp.
  159--168.

\bibitem{ref_xilinx}
B.~Gaide, D.~Gaitonde, C.~Ravishankar, and T.~Bauer, ``Xilinx adaptive compute
  acceleration platform: Versaltm architecture,'' in \emph{Proceedings of the
  2019 ACM/SIGDA International Symposium on Field-Programmable Gate Arrays},
  2019, pp. 84--93.

\bibitem{ref_npn}
X.~Zhou, L.~Wang, and A.~Mishchenko, ``Fast exact npn classification by
  co-designing canonical form and its computation algorithm,'' \emph{IEEE
  Transactions on Computers}, vol.~69, no.~9, pp. 1293--1307, 2020.

\bibitem{ref_dsd_map}
A.~Mishchenko, R.~Brayton, W.~Feng, and J.~Greene, ``Technology mapping into
  general programmable cells,'' in \emph{Proceedings of the 2015 ACM/SIGDA
  International Symposium on Field-Programmable Gate Arrays}, 2015, pp. 70--73.

\bibitem{ref_cegar_qsat}
M.~Janota, W.~Klieber, J.~Marques-Silva, and E.~Clarke, ``Solving qbf with
  counterexample guided refinement,'' \emph{Artificial Intelligence}, vol. 234,
  pp. 1--25, 2016.

\bibitem{ref_hyperopt}
J.~Bergstra, D.~Yamins, and D.~Cox, ``Making a science of model search:
  Hyperparameter optimization in hundreds of dimensions for vision
  architectures,'' in \emph{International conference on machine
  learning}.\hskip 1em plus 0.5em minus 0.4em\relax PMLR, 2013, pp. 115--123.

\bibitem{ref_if}
A.~Mishchenko, S.~Cho, S.~Chatterjee, and R.~Brayton, ``Combinational and
  sequential mapping with priority cuts,'' in \emph{2007 IEEE/ACM International
  Conference on Computer-Aided Design}.\hskip 1em plus 0.5em minus 0.4em\relax
  IEEE, 2007, pp. 354--361.

\bibitem{ref_dsd}
A.~Mishchenko and R.~Brayton, ``Faster logic manipulation for large designs,''
  in \emph{Proc. IWLS}, vol.~13, 2013.

\bibitem{ref_mfs}
A.~Mishchenko, R.~Brayton, J.-H.~R. Jiang, and S.~Jang, ``Scalable
  don't-care-based logic optimization and resynthesis,'' \emph{ACM Transactions
  on Reconfigurable Technology and Systems (TRETS)}, vol.~4, no.~4, pp. 1--23,
  2011.

\bibitem{ref_coffe}
\BIBentryALTinterwordspacing
S.~Yazdanshenas and V.~Betz, ``Coffe 2: Automatic modelling and optimization of
  complex and heterogeneous fpga architectures,'' \emph{ACM Trans.
  Reconfigurable Technol. Syst.}, vol.~12, no.~1, jan 2019. [Online].
  Available: \url{https://doi.org/10.1145/3301298}
\BIBentrySTDinterwordspacing

\end{thebibliography}

%\bibliographystyle{IEEEtran}

\end{document}